%% file: Bcondensate.tex
\documentclass[aps,prl,reprint,superscriptaddress,raggedbottom,nobibnotes,nofootinbib,a4paper,showpacs,twocolumn,lengthcheck]{revtex4-1}
\pdfoutput=1
\usepackage{amsfonts,amssymb,epsfig,verbatim,mathbbol,amstext,amsmath,psfrag}
\usepackage[bookmarks,
                 bookmarksopen = true,
                 bookmarksnumbered = true,
                 linktocpage,
                 colorlinks = true,
                 linkcolor = blue,
                 urlcolor  = blue,
                 citecolor = blue,
                 anchorcolor = green,
                 hyperindex = true,
                 hyperfigures]
		{hyperref}
\usepackage{slashed}
\usepackage{dsfont}
\usepackage{float}
\usepackage{multirow}

\newcommand{\be}{\begin{equation}}
\newcommand{\ee}{\end{equation}}

\newcommand{\Z}{\mathcal{Z}}
\newcommand{\D}{\mathcal{D}}

\begin{document}

\title{QCD quark condensate in external magnetic fields}

\author{G.~S.~Bali}
\affiliation{Institute for Theoretical Physics, Universit\"at Regensburg, D-93040 Regensburg, Germany.}
\affiliation{Department of Theoretical Physics, Tata Institute of Fundamental Research, 
Homi Bhabha Road, Mumbai 400005, India.}
\author{F.~Bruckmann}
\affiliation{Institute for Theoretical Physics, Universit\"at Regensburg, D-93040 Regensburg, Germany.}
\author{G.~Endr\H{o}di}
\affiliation{Institute for Theoretical Physics, Universit\"at Regensburg, D-93040 Regensburg, Germany.}
\author{Z.~Fodor}
\affiliation{Department of Physics, Bergische Universit\"at Wuppertal, D-42119, Germany.}
\affiliation{Institute for Theoretical Physics, E\"otv\"os University, H-1117 Budapest, Hungary.}
\affiliation{J\"ulich Supercomputing Centre, Forschungszentrum J\"ulich, D-52425 J\"ulich, Germany. }
\author{S.~D.~Katz}
\affiliation{Institute for Theoretical Physics, E\"otv\"os University, H-1117 Budapest, Hungary.}
\author{A.~Sch\"afer}
\affiliation{Institute for Theoretical Physics, Universit\"at Regensburg, D-93040 Regensburg, Germany.}

\begin{abstract}
We present a comprehensive analysis of the light condensates in QCD with 1+1+1 sea quark flavors (with mass-degenerate light quarks of different electric charges) at zero and nonzero temperatures of up to $190 \textmd{ MeV}$ and external magnetic fields $B<1\,\mathrm{GeV}^2/e$. We employ stout smeared staggered fermions with physical quark masses and extrapolate the results to the continuum limit. At low temperatures we confirm the magnetic catalysis scenario predicted by many model calculations while around the crossover the condensate develops a complex dependence on the external magnetic field, resulting in a decrease of the transition temperature. 
\end{abstract}



\maketitle

\emph{Introduction}.---Strong (electro)magnetic fields prominently feature in various physical systems. They play an essential role in cosmology, where magnetic fields of $10^{14}\textmd{ T}$ and $10^{19}\textmd{ T}$ may have been present~\cite{Vachaspati:1991nm,Enqvist:1993kf} during the strong and electroweak phase transitions of the universe, respectively. Magnetic fields with strengths up to $B\sim 10^{14-16} \textmd{ T}$ ($\sqrt{eB}\sim 0.1-1.0$~GeV)
are also generated in non-central heavy ion collisions~\cite{Skokov:2009qp,Voronyuk:2011jd,Bzdak:2011yy,Deng:2012pc} at the Relativistic Heavy Ion Collider (RHIC) or the Large Hadron Collider (LHC). Furthermore, for certain classes of neutron stars like magnetars, magnetic fields of the order of $10^{10} \textmd{ T}$ have been deduced~\cite{Duncan:1992hi}. In addition to this phenomenological relevance, external (electro)magnetic fields can be used to probe the dynamics of strongly interacting matter, i.e.\ the vacuum structure of Quantum Chromodynamics (QCD). 

One of the most important aspects of QCD is chiral symmetry breaking. At zero quark masses the chiral condensate $\bar\psi\psi$
is an order parameter.
It vanishes at high temperatures where chiral symmetry is restored but develops
a nonzero expectation value in the hadronic phase.
In nature, quark masses are nonzero and the corresponding quark condensates, though only approximate order parameters, still exhibit this characteristic behavior around the transition temperature between the hadronic and the quark-gluon plasma phases.
Lattice simulations revealed that for physical quark masses this transition is an analytic crossover~\cite{Aoki:2006we}, leading to 
a transition temperature $T_c$ which may depend on the observable used for its definition.

\enlargethispage{\baselineskip}

In the response of QCD to external magnetic fields, `magnetic catalysis' refers to an increase of the condensate with $B$. 
This implies a $B$-dependence of $T_c$ as well. Almost all low-energy
models and approximations to QCD~\cite{Gusynin:1995nb,Shushpanov:1997sf,Agasian:1999sx,Agasian:2001hv,Cohen:2007bt,Andersen:2012dz,Andersen:2012zc,Klevansky:1989vi,Menezes:2009uc,Gatto:2010pt,Gatto:2010qs,Kashiwa:2011js,Andersen:2011ip,Avancini:2012ee,Fukushima:2012xw,Mizher:2010zb,Gatto:2010pt,Andersen:2012bq,Kanemura:1997vi,Klimenko:1992ch,Alexandre:2000yf,Scherer:2012nn,Johnson:2008vna,Preis:2010cq,Gusynin:1995nb,Shushpanov:1997sf,Agasian:1999sx,Cohen:2007bt,Agasian:2001hv,Mizher:2011wd,Gasser:1986vb,Gasser:1987ah,Gerber:1988tt} as well as lattice simulations in quenched theories \cite{Buividovich:2008wf,Braguta:2010ej} and at larger than physical pion masses in $N_f=2$ QCD \cite{D'Elia:2010nq,D'Elia:2011zu} and in the $N_f=4$ SU(2) theory \cite{Ilgenfritz:2012fw} found $\bar\psi\psi(B)$ and $T_c(B)$ to increase with $B$. 
Exceptions in this respect with a decreasing $T_c(B)$ function are the results obtained within 
two-flavor chiral perturbation theory~\cite{Agasian:2008tb}, in the linear sigma model without vacuum corrections~\cite{Fraga:2008um} and in the bag model~\cite{Fraga:2012fs}.

In contrast to the majority of the above results, our large-scale study of QCD in external magnetic fields with physical pion mass $M_\pi=135 \textmd{ MeV}$ and results extrapolated to the continuum limit~\cite{Bali:2011qj} has revealed the transition temperature to \emph{decrease} as a function of the external magnetic field. This applies to the $T_c$'s 
defined from the quark condensate, the strange quark number susceptibility and the chiral susceptibility. In particular, we found the condensate to depend on $B$ in a \emph{non-monotonous} way in the crossover region. 

In Ref.~\cite{Bali:2011qj} we have already pointed out two rationales why former lattice studies are at variance with these recent findings: coarser lattices and larger quark masses. Obviously, it is also very important to
address the differences between our QCD results and many model and chiral perturbation theory ($\chi$PT) predictions,
especially since the latter methods can be used to investigate regions that are not easily accessible to lattice simulations, e.g., QCD at a non-vanishing baryon density. 

In this Letter, we present a detailed analysis of the dependence of the light quark condensates
on $B$ and on the temperature $T$, based on the $T>0$ simulations described in~\cite{Bali:2011qj} and new
simulations at $T=0$.
The data --- all continuum extrapolated --- are presented in ways that will enable to refine model assumptions and parameters. We also perform a first comparison to $\chi$PT and to Polyakov-Nambu-Jona-Lasinio (PNJL) model predictions. We aim at a better understanding of the physical mechanisms behind the differences. This in turn should be of phenomenological relevance.

Below we introduce our notations and the simulation setup.
We then present our results and compare them to $\chi$PT and PNJL predictions, both at
zero and nonzero temperatures.

\emph{Condensate on the lattice at nonzero $B$}.---We study QCD coupled to a constant external magnetic field $B$, pointing in the positive $z$ direction. Such a field can be implemented by multiplying the $U\in\mathrm{SU}(3)$ links of the lattice by complex phases. The specific choice of these phases and our setup are detailed in Ref.~\cite{Bali:2011qj}. 
The external field couples only to the quark electric charges $q_f$ with $f$ labelling the different flavors. Thus, the magnetic field only appears in combinations $q_fB$. 

In a finite periodic volume, the magnetic flux is quantized~\cite{'tHooft:1979uj,AlHashimi:2008hr}. This quantization on a lattice with spacing $a$ amounts to,
\be
(N_sa)^2 \cdot q_d B = 2\pi N_b, \quad\quad N_b\in\mathds{Z},
\label{eq:Bquant}
\ee
where the smallest quark charge (that of the down quark), $|q_d|=e/3$ enters, with $e>0$ being the elementary charge.
Here $N_s$ is the number of lattice sites in a spatial direction (our lattices are symmetric in space). Similarly, $N_t$ counts the lattice points in the temporal direction. The spatial volume of the system is given by $V=(N_sa)^3$ and the temperature is related to the inverse temporal extent of the lattice as $T=(N_ta)^{-1}$. 

The quark condensate can be derived from the partition function, which in the staggered formulation of QCD with three flavors ($f=u,d,s$) is given by the functional integral,
\be
\Z = \int \D U \,e^{-\beta S_g} \prod_{f=u,d,s} \left[ \det M(U,q_fB,m_f)\right]^{1/4},
\label{eq:partfunc}
\ee
where $\beta\equiv 6/g^2$ is the inverse gauge coupling, $S_g$ the gauge action and $M(U,qB,m) = \slashed D(U,qB) + m\mathds{1}$ the fermion matrix. For $S_g$ we use the tree-level improved Symanzik action, while in the fermionic sector we employ a stout smeared staggered Dirac operator $\slashed{D}$. The details of the lattice action can be found in Refs.~\cite{Aoki:2005vt,Bali:2011qj}. The lattice sizes range from $24^3\times 32$ to $40^3\times48$ for the zero temperature simulations, while at non-vanishing $T$ we investigate $24^3\times 6$, $24^3\times 8$ and $28^3\times 10$ lattices. 
We set the quark masses to their physical values, with mass-degenerate light quarks: $m_u=m_d\equiv m_{ud}$. The electric charges of the quarks are $q_d=q_s=-q_u/2=-e/3$, therefore we need to treat each flavor separately. The line of constant physics (LCP) $[m_{ud}(\beta),m_s(\beta)]$ was determined by fixing the ratios $M_\pi/f_K$ and $M_K/f_K$ to the experimental values. The lattice spacing $a(\beta)$ is defined by keeping $f_K=f_K^{\mathrm{lat}}(\beta)/a(\beta)$ fixed, for details see Ref.~\cite{Borsanyi:2010cj}.
At $T=0$ the continuum limit $a\to0$ corresponds to $\beta\to\infty$. At nonzero temperature, it is convenient to define the continuum limit as $N_t\to\infty$, keeping $T$ fixed.

\begin{figure}[t!]
\includegraphics*[width=8.5cm]{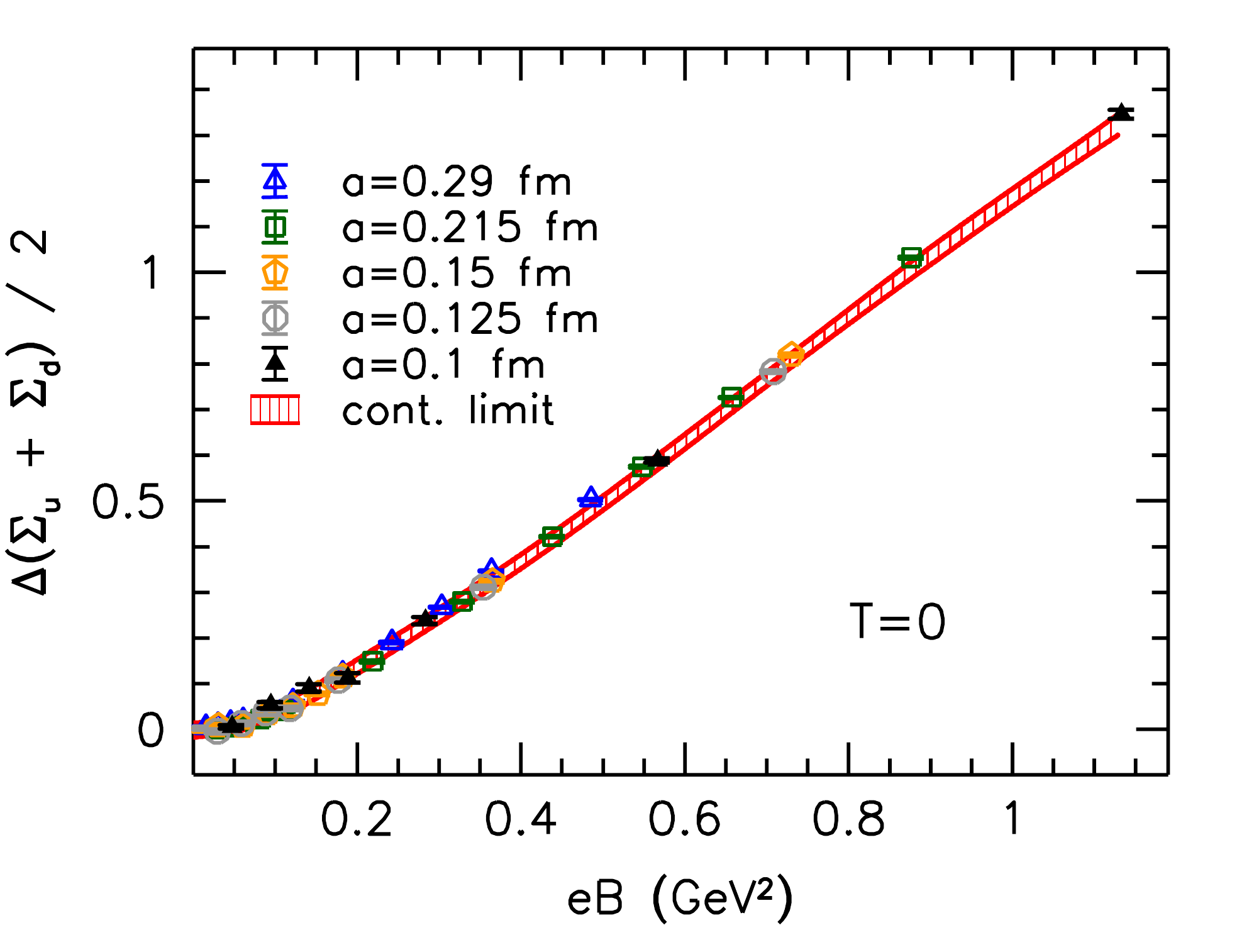}
\vspace*{-0.4cm}
\caption{The change of the renormalized condensate due to the magnetic field at $T=0$ as measured on five lattice spacings and the continuum limit.}
\label{fig:contfitT0}
\end{figure}

The quark condensate is defined as the derivative of $\ln\Z$ with respect to the lattice mass parameter
\be
\bar\psi\psi_f(B,T) \equiv \frac{T}{V} \frac{\partial \ln\Z(B,T)}{\partial m_f} .
\label{eq:pbpdef}
\ee
To carry out the continuum limit, the lattice condensate $\bar\psi\psi$ needs to be renormalized since it contains additive (for $m>0$) and multiplicative divergences. These cancel~\cite{Bali:2011qj} in the following combination,
\be
\Sigma_{u,d}(B,T) = \frac{2m_{ud}}{M_\pi^2 F^2} \left[ \bar\psi\psi_{u,d}(B,T)- \bar\psi\psi_{u,d}(0,0) \right] + 1, 
\label{eq:pbpren}
\ee
where, to obtain a dimensionless quantity, we divided by the combination $M_\pi^2 F^2$ which contains the zero-field pion mass $M_\pi=135 \textmd{ MeV}$ and (the chiral limit of the) pion decay constant $F=86 \textmd{ MeV}$~\cite{Colangelo:2003hf}. This specific combination enters the Gell-Mann-Oakes-Renner relation,
\be
2m_{ud} \cdot \bar\psi\psi(0,0) = M_\pi^2 F^2+\cdots.
\ee
Note that the normalization in definition~(\ref{eq:pbpren}) can easily be converted into the slightly different one employed in former studies by the Budapest-Wuppertal collaboration (e.g. Refs.~\cite{Aoki:2006we,Borsanyi:2010bp}) and in Ref.~\cite{Bali:2011qj}. We define the change of the condensate due to the magnetic field as
\be
\Delta\Sigma_{u,d}(B,T) = \Sigma_{u,d}(B,T)-\Sigma_{u,d}(0,T).
\label{eq:deltapbp}
\ee
Note that the $\bar\psi\psi(0,0)$ term cancels from this difference.
In our normalization, Eq.~(\ref{eq:deltapbp}) defines the change of the condensate caused by a nonzero $B$, in units of the chiral condensate at $B=0$ and $T=0$. This normalization will be advantageous when comparing the lattice results to $\chi PT$ and model predictions, which are usually given in units of $\bar\psi\psi(0,0)$. The $+1$ is included in Eq.~(\ref{eq:pbpren}) so that the chiral limit of the condensate is fixed to 1 at $T=B=0$, and approaches 0 as $T\to\infty$. At nonzero quark mass $\Sigma_{u,d}$ will still start from 1 at $T=B=0$. 
At very high temperatures, however, it is well known from the
free case~\cite{PhysRevD.9.3320,PhysRevD.9.3357} that the condensate receives a contribution
$\sim mT^2$. This term is negligible for the temperatures under study and it cancels exactly from $\Delta\Sigma_{u,d}$.

\emph{Results}.---In Fig.~\ref{fig:contfitT0} we display the renormalized difference $\Delta(\Sigma_u+\Sigma_d)/2$ as a function of $B$ at $T=0$, for five different lattice spacings. We carry out the continuum limit by fitting the results to a lattice spacing-dependent spline function (for a similar fit in two dimensions see~\cite{Endrodi:2010ai}). This function is defined on a set of points and is parameterized by two values at each such node, in the form $c_k + a^2 d_k$, to reflect the $a^2$-scaling of our action. The parameters $c_k$ and $d_k$ are obtained by minimizing the corresponding $\chi^2$. The systematic error of the $a\to0$ limit is determined by varying the node positions. 
We find that lattice discretization errors become large at high magnetic fields due to saturation of the lattice magnetic flux~\cite{Bali:2011qj}, therefore we only include points with $N_b/N_s^2<0.1$.
In Fig.~\ref{fig:contfitT0} we also show the continuum limit of the difference $\Delta(\Sigma_u+\Sigma_d)/2$.

\begin{figure}[t]
\includegraphics*[width=8.5cm]{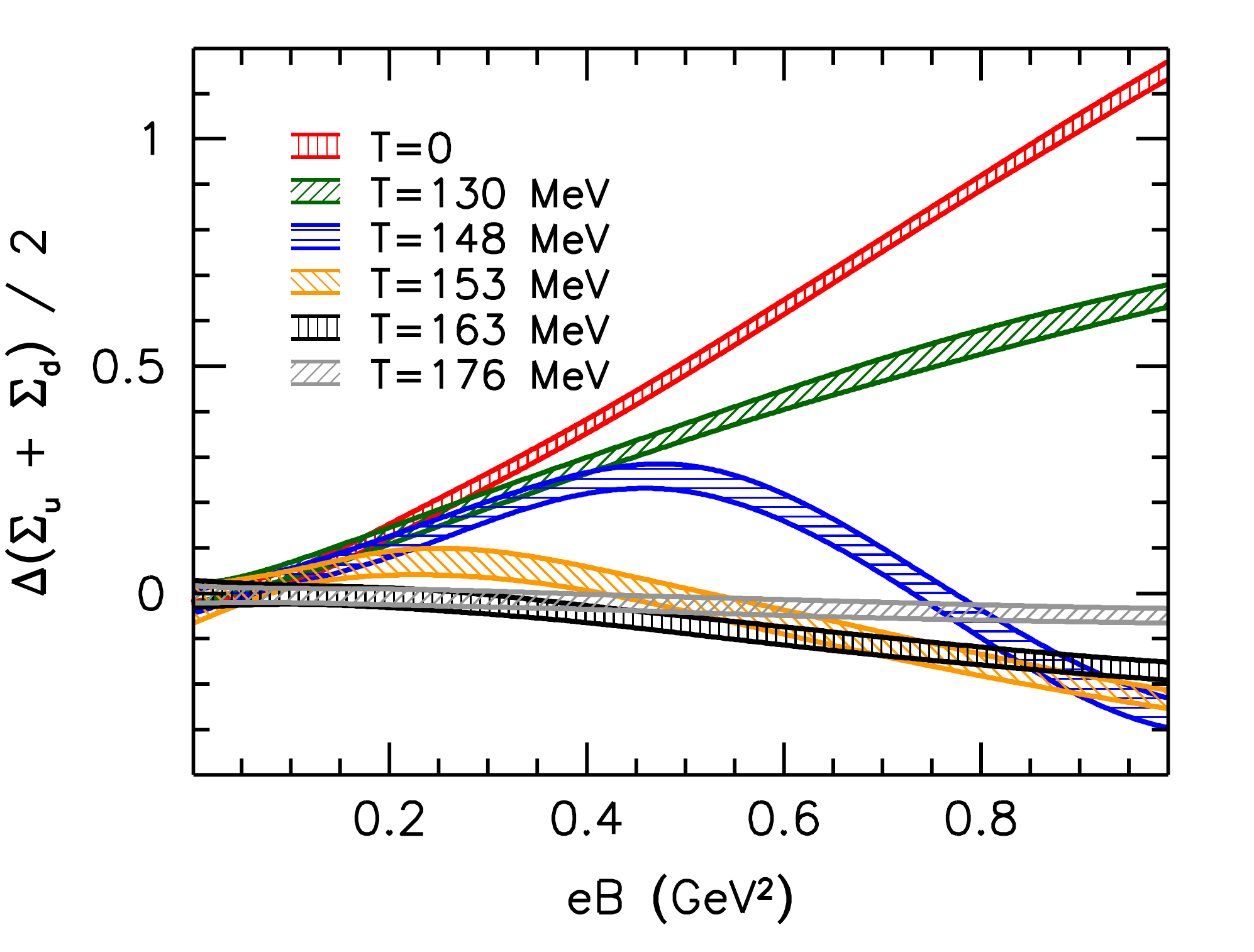}
\vspace*{-0.4cm}
\caption{Continuum extrapolated lattice results for the change
of the condensate as a function of $B$, at six different temperatures.}
\label{fig:pbpBT}
\end{figure}

Next, we address the condensate at nonzero temperature, carrying out a similar continuum extrapolation for $\Delta\Sigma$ as at $T=0$, using three lattice spacings with $N_t=6$, $8$ and $10$. The increase of the difference $\Delta\Sigma(B)$ is qualitatively similar for zero and nonzero temperatures in $\chi$PT and in the PNJL model (see below). In QCD, however, the situation is quite different: in Fig.~\ref{fig:pbpBT} we plot the continuum extrapolated lattice results for $\Delta(\Sigma_u+\Sigma_d)/2$ as functions of $B$ for several temperatures, ranging from $T=0$ up to $T=176 \textmd{ MeV}$. Note that the transition temperature varies from $T_c(eB=0)\approx 158 \textmd{ MeV}$ down to $T_c(0.9\textmd{ GeV}^2)\approx 138 \textmd{ MeV}$~\cite{Bali:2011qj}.
The increasing behavior of $\Delta\Sigma(B)$ at low temperatures ($T\le 130 \textmd{ MeV}$) corresponding to magnetic catalysis continuously transforms into a hump-like structure in the crossover region ($T=148\textmd{ MeV}, 153 \textmd{ MeV}$) and then on to a monotonously decreasing dependence ($T\ge 163\textmd{ MeV}$). We remark that --- although in the high temperature limit the condensate and its dependence on $B$ are suppressed --- at $T\gtrsim 190 \textmd{ MeV}$ $\Delta\Sigma(B)$ again starts to increase.
Furthermore, we note that the strange condensate $\Delta\Sigma_s$ (with a definition similar to that in Eq.~(\ref{eq:pbpren})) does not exhibit this complex dependence on $B$ and $T$ but simply increases with growing $B$ for all temperatures. This shows that the partly decreasing behavior near the crossover region only appears for quark masses below a certain threshold $m_{\rm thr}$, inbetween the physical light and strange quark masses, $m_{ud}<m_{\rm thr}<m_s$.

\begin{figure}[t]
\includegraphics*[width=8.5cm]{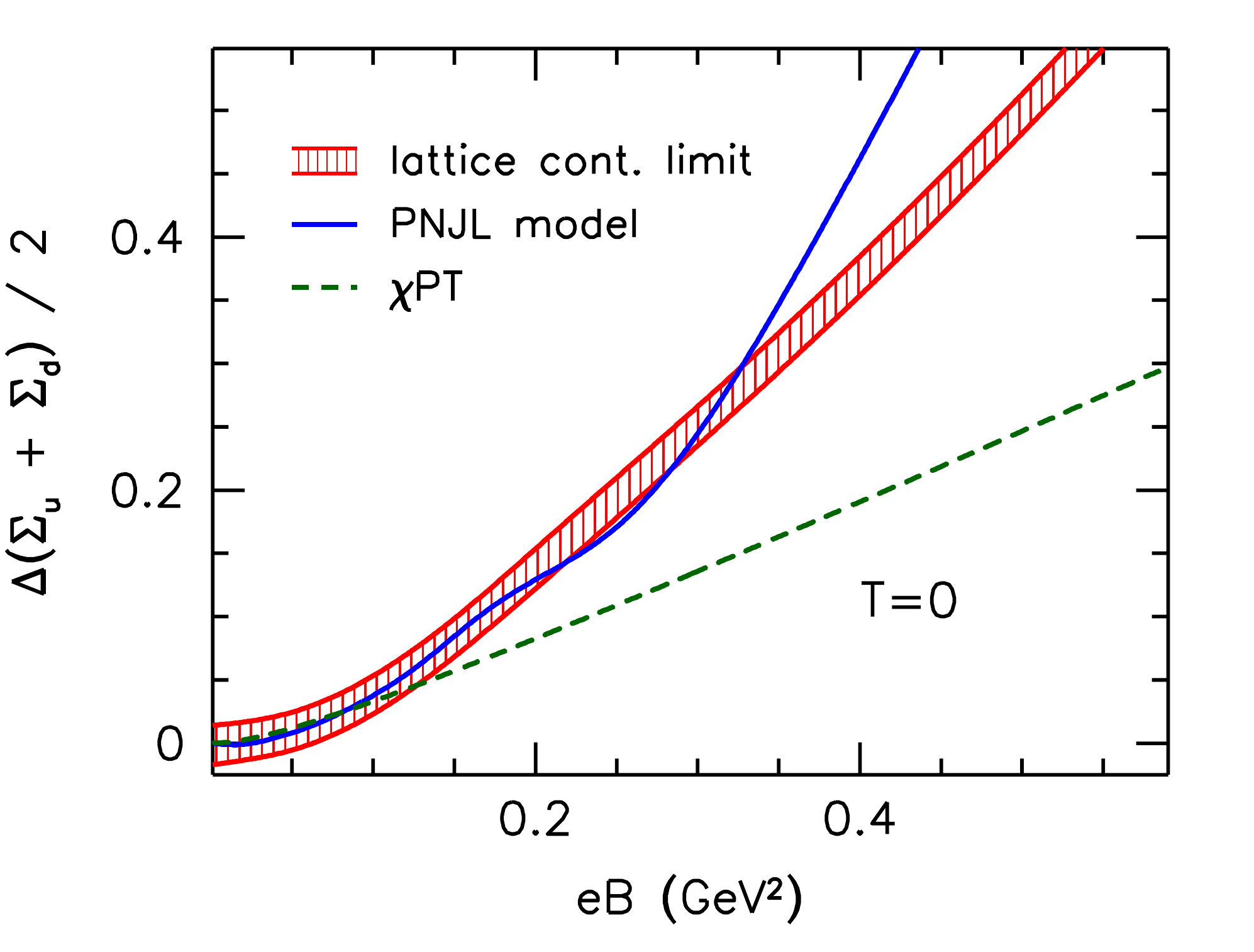}
\vspace*{-0.4cm}
\caption{Comparison of the continuum limit of the change of the condensate to the $\chi$PT~\protect\cite{Cohen:2007bt,Andersen:2012dz,Andersen:2012zc,Andersen} and the (P)NJL model \protect\cite{Gatto:2010pt,Ruggieri} predictions.}
\label{fig:cmpT0}
\end{figure}

\emph{Comparison to effective theories/models}.---In Fig.~\ref{fig:cmpT0} we compare our zero temperature QCD result for $\Delta(\Sigma_u+\Sigma_d)/2$
as a function of $B$ to the $\chi$PT prediction~\cite{Cohen:2007bt,Andersen:2012dz,Andersen:2012zc,Andersen} and to that of the PNJL model~\cite{Gatto:2010pt,Ruggieri}, both at physical pion mass. We see that the $\chi$PT prediction describes the lattice results well up to $eB=0.1 \textmd{ GeV}^2$, while the PNJL model works quantitatively well up to $eB = 0.3 \textmd{ GeV}^2$.
Note that, since the Polyakov loop at zero temperature vanishes, in the limit $T\to0$ the PNJL model becomes indistinguishable from the NJL model with the same couplings.

\begin{figure}[t]
\includegraphics*[width=8.5cm]{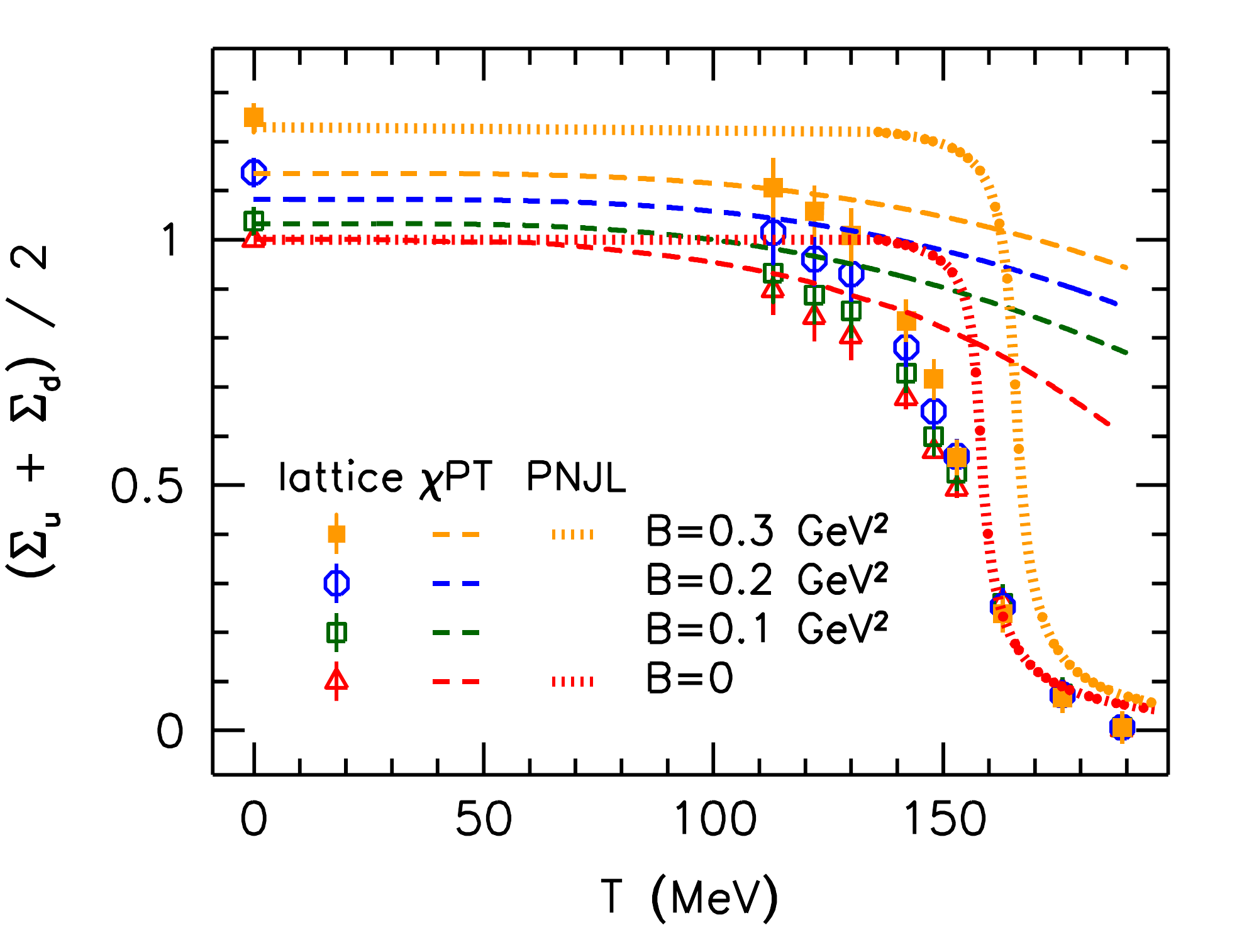}
\vspace*{-0.4cm}
\caption{Comparison of the continuum extrapolated lattice results (points) to $\chi$PT \protect\cite{Andersen:2012dz,Andersen:2012zc,Andersen} (dashed lines) and the PNJL model \protect\cite{Gatto:2010pt,Ruggieri} (dotted lines) at different magnetic fields.}
\label{fig:cmpT}
\end{figure}

In Fig.~\ref{fig:cmpT}, the condensate Eq.~(\ref{eq:pbpren}) as a function of $T$ is compared to $\chi$PT and to the PNJL model for different magnetic fields. 
At $B=0$ we use the continuum extrapolation for the condensate presented in Ref.~\cite{Borsanyi:2010bp} (where lattices up to $N_t=16$ were employed), and complement this with the differences $\Delta\Sigma(B)$ shown in Fig.~\ref{fig:pbpBT}.
In addition to the continuum extrapolated lattice data we plot the $\chi$PT curves for $B=0$~\cite{Gerber:1988tt} and for $B>0$~\cite{Andersen:2012dz,Andersen:2012zc,Andersen}, together with the PNJL model predictions~\cite{Gatto:2010pt,Ruggieri}. 
The results indicate that $\chi$PT is reliable for small temperatures and small magnetic fields, $eB\lesssim 0.1 \textmd{ GeV}^2$, $T\lesssim 100 \textmd{ MeV}$.
(We remark that the inclusion of the hadron resonance gas contribution to the condensate in $\chi$PT~\cite{Gerber:1988tt} improves the agreement with lattice results, as was shown at $B=0$ in Ref.~\cite{Borsanyi:2010bp}. One would expect a similar improvement at $B>0$.)
Since the PNJL model condensate is calculated using a Polyakov loop effective potential that was obtained from $N_f=2$ lattice results~\cite{Gatto:2010pt}, differences between the model and our $N_f=1+1+1$ results at $T>0$ are expected to be large, as both the transition temperature and the transition strength (the slope of the condensate at $T_c$) strongly depend on the number of flavors. To enable a comparison, we linearly rescaled the temperature axis (only for the PNJL curves) to match our lattice inflection point at $B=0$. Nevertheless, the $B$-dependence of the condensate for the PNJL model also reveals qualitative differences in comparison to the QCD results.

\begin{figure}[t]
\includegraphics*[width=8.5cm]{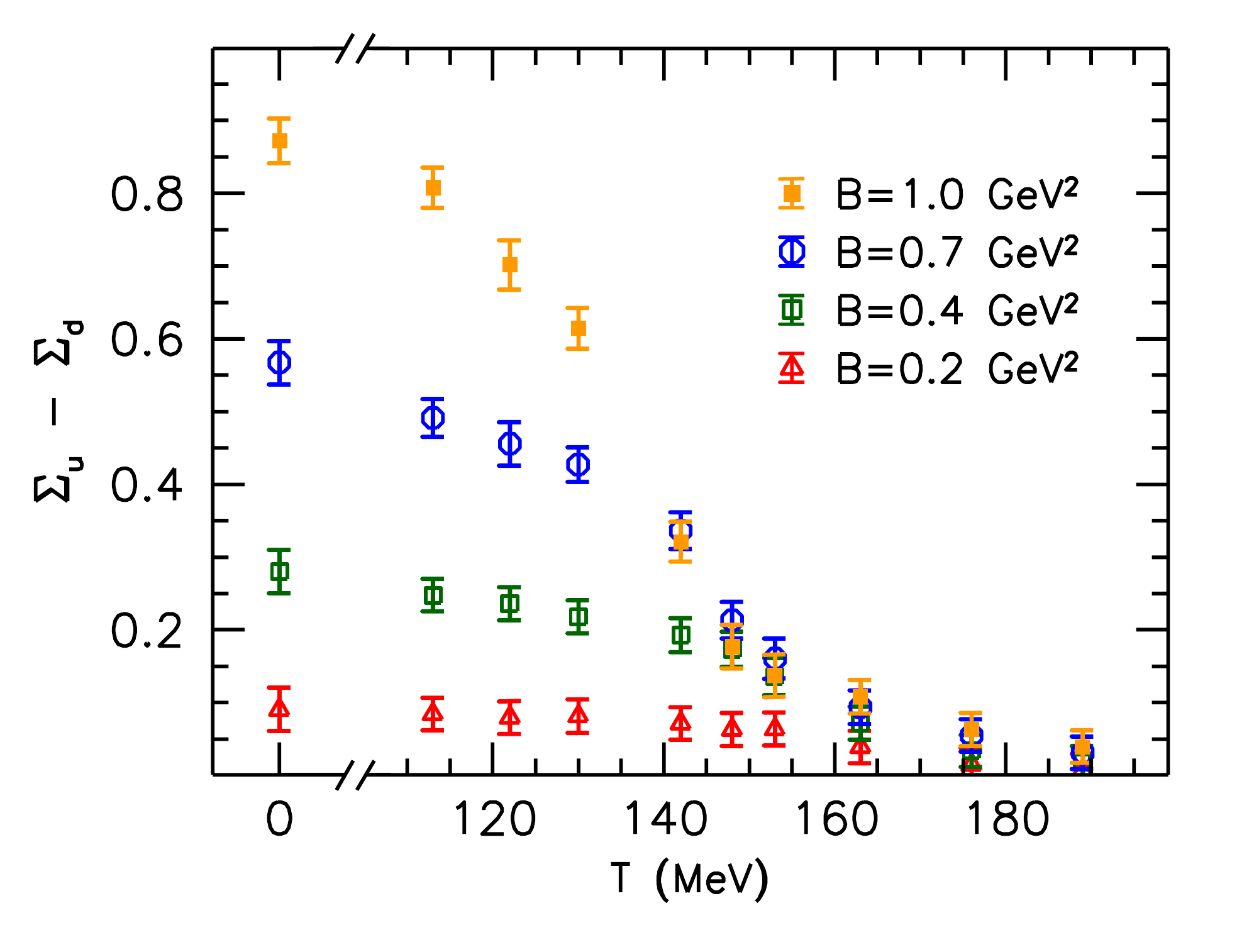}
\vspace*{-0.4cm}
\caption{Continuum extrapolated results for the difference of the up and the down quark condensates. }
\label{fig:uminusd}
\end{figure}

Finally, in Fig.~\ref{fig:uminusd} we plot $\Delta\Sigma_u-\Delta\Sigma_d = \Sigma_u-\Sigma_d$ as a function of the temperature for several magnetic field strengths. At zero magnetic field isospin symmetry is exact since we employed mass-degenerate light quarks. As $B$ increases, due to the difference between the electric charges, $\Sigma_u-\Sigma_d$ develops a temperature-dependence similar to that of $(\Sigma_u+\Sigma_d)/2$, see Fig.~\ref{fig:cmpT}. The results for $\Sigma_u \pm \Sigma_d$ are also listed in Table~\ref{tab1}.

\emph{Summary}.---We determined the QCD light quark condensates at nonzero external magnetic field strengths for physical quark masses in the continuum limit. Our results are in quantitative agreement with chiral perturbation theory and PNJL model predictions for small magnetic fields and at small temperatures. 
Note that the constants within these parameterizations have not been adjusted to our data but were
taken from the literature where they have been obtained at vanishing magnetic field.
Unsurprisingly, $\chi$PT fails in regions where pions cease to be the essential low energy degrees of freedom. While in the hadronic phase low energy models qualitatively reproduce the $B$-dependence of the lattice data, they miss an important feature which becomes dominant for light quark masses and for temperatures around $T_c$, see Fig.~\ref{fig:pbpBT}. 
Clearly, the coupling between the magnetic field and the gauge background is enhanced near the chiral limit: the smaller the quark mass, the more the fluctuations of the gauge field influence the quark determinant. Thus, for light quarks the indirect interaction between the gluonic degrees of freedom and the external field becomes more important. A possibility to separate this indirect effect would be to consider the sea and valence contributions to the condensate, as was performed in Ref.~\cite{D'Elia:2011zu}, which we plan to discuss in a forthcoming study.

\emph{Acknowledgments}.---
We thank J.O.\ Andersen, T.\ Cohen, E.\ Fraga, T.\ Kov\'acs, M.\ Ruggieri and K.\ Szab\'o
for useful discussions.
Computations were carried out on the GPU cluster~\cite{Egri:2006zm} at E\"otv\"os University
Budapest and the BlueGene/P of FZ J\"ulich.
This work was supported by 
the DFG (SFB/TR 55 and BR 2872/4-2) and the EU
(ITN STRONGnet 238353 and ERC Grant 208740).

\begin{center}
\begin{table}[ht!]
\fontsize{9}{8.5}\selectfont
\hspace*{-0.1cm}\mbox{
\input{table}

}
\caption{Continuum extrapolated lattice results for the light condensates, as functions of $T$ and $eB$. Columns labeled `$+/2$' contain the average $(\Sigma_u+\Sigma_d)/2$, while those with `$-$' contain the difference $\Sigma_u-\Sigma_d$. Note that the uncertainty of the lattice scale gives rise to errors of about $2\%$ in the temperatures.
\label{tab1}
}
\end{table}
\end{center}

\bibliographystyle{apsrev}
\bibliography{Bcondensate}

\end{document}

%% file: table.tex
\begin{tabular}{|c||c|c||c|c||c|c|}
\hline
\multirow{2}{*}{$T (\textmd{MeV})$} 
& \multicolumn{2}{|c||}{$eB=0$} 
& \multicolumn{2}{|c||}{$eB=0.2 \textmd{ GeV}^2$}
& \multicolumn{2}{|c|}{$eB=0.4 \textmd{ GeV}^2$} \\
\cline{2-7}
& $+/2$ & $-$
& $+/2$ & $-$ 
& $+/2$ & $-$ \\ 
\hline\hline
0 & 1 & 0 & 1.14(2) & 0.09(2) & 1.37(2) & 0.28(2) \\ 
113 & 0.90(4) & 0 & 1.01(6) & 0.08(2) & 1.21(5) & 0.25(2) \\ 
122 & 0.84(4) & 0 & 0.96(5) & 0.08(2) & 1.17(5) & 0.24(3) \\ 
130 & 0.80(4) & 0 & 0.93(5) & 0.08(3) & 1.09(5) & 0.22(2) \\ 
142 & 0.68(2) & 0 & 0.78(3) & 0.07(2) & 0.89(4) & 0.19(3) \\ 
148 & 0.57(1) & 0 & 0.65(3) & 0.06(2) & 0.76(6) & 0.17(3) \\ 
153 & 0.49(1) & 0 & 0.56(3) & 0.06(2) & 0.53(3) & 0.14(3) \\ 
163 & 0.26(1) & 0 & 0.25(3) & 0.04(2) & 0.22(3) & 0.07(3) \\ 
176 & 0.08(1) & 0 & 0.07(3) & 0.01(2) & 0.06(3) & 0.03(2) \\ 
189 & 0.00(1) & 0 & 0.01(3) & 0.01(2) & 0.00(3) & 0.02(2) \\ 
\hline\hline
\multirow{2}{*}{$T (\textmd{MeV})$} 
& \multicolumn{2}{|c||}{$eB=0.6 \textmd{ GeV}^2$} 
& \multicolumn{2}{|c||}{$eB=0.8 \textmd{ GeV}^2$}
& \multicolumn{2}{|c|}{$eB=1.0 \textmd{ GeV}^2$} \\
\cline{2-7}
& $+/2$ & $-$
& $+/2$ & $-$ 
& $+/2$ & $-$ \\ 
\hline\hline
0 & 1.63(3) & 0.47(3) & 1.90(3) & 0.67(3) & 2.16(3) & 0.87(3) \\ 
113 & 1.48(6) & 0.41(3) & 1.73(6) & 0.58(3) & 1.95(4) & 0.81(3) \\ 
122 & 1.40(5) & 0.38(3) & 1.63(5) & 0.53(3) & 1.86(6) & 0.70(3) \\ 
130 & 1.23(5) & 0.36(3) & 1.36(5) & 0.49(3) & 1.46(5) & 0.61(3) \\ 
142 & 0.94(4) & 0.30(3) & 0.85(4) & 0.35(3) & 0.68(4) & 0.32(3) \\ 
148 & 0.66(5) & 0.22(3) & 0.50(4) & 0.20(3) & 0.38(4) & 0.18(3) \\ 
153 & 0.43(3) & 0.17(3) & 0.34(3) & 0.15(3) & 0.26(3) & 0.14(3) \\ 
163 & 0.17(3) & 0.09(3) & 0.12(3) & 0.10(3) & 0.09(3) & 0.11(3) \\ 
176 & 0.05(3) & 0.05(2) & 0.04(3) & 0.06(2) & 0.03(3) & 0.06(2) \\ 
189 & -0.00(3) & 0.03(2) & -0.01(3) & 0.03(2) & -0.01(3) & 0.04(2) \\ 
\hline
\end{tabular}